\documentclass[11pt,letterpaper]{article}

\usepackage{jheppub}
\usepackage{xcolor}
\usepackage{graphicx}
\usepackage{xspace}
\usepackage{wrapfig,enumerate,slashed}

\usepackage{footmisc}
\usepackage{amsmath}
\usepackage{wasysym} 
\usepackage{graphicx}
\usepackage{color}
\usepackage{comment}
\usepackage{hyperref}
\usepackage{subfigure}

\newcommand{\be}{\begin{eqnarray}}
\newcommand{\ee}{\end{eqnarray}}

\newcommand{\bee}{\begin{eqnarray}}
\newcommand{\eee}{\end{eqnarray}}
\newcommand{\beeq}{\begin{equation}}
\newcommand{\eeeq}{\end{equation}}

\newcommand{\SU}{\text{SU}}

\newcommand{\rL}{\ensuremath{\sqrt{\mathcal{L}}}\xspace}

\bibstyle{plain}
\newcommand\allFontSize{\small}

\newcommand\detailsSize{\allFontSize}
{\begin{myquote}\vspace{-0.2cm}\detailsSize}{\end{myquote}\vspace{-0.2cm}}

\definecolor{DarkGray}{rgb}{0.4,0.42,0.45}
\definecolor{LightGray}{rgb}{0.97,0.98,0.98}

\usepackage{tabularx}

\newcommand{\pt}        {\ensuremath{p_\mathrm{T}}\xspace}

\newcommand{\nECF}[2][]
{
  \ifthenelse{\isempty{#1}}
  {\ensuremath{e_{#2}}\xspace}
  {\ensuremath{e_{#2}^{(#1)}}\xspace}
}

\newcommand{\bq}    {\ensuremath{b}\xspace}

\newcommand{\ttbar} {\ensuremath{t\overline t}\xspace}
\newcommand{\fourtop} {\ensuremath{t\overline{t}t\overline{t}}\xspace}

\newcommand{\tWj} {\ensuremath{tW\!j}\xspace}
\newcommand{\tZj} {\ensuremath{tZ\!j}\xspace}
\newcommand{\tWZ} {\ensuremath{tW\!Z}\xspace}

\newcommand{\ttZ} {\ensuremath{t\overline{t}Z}\xspace}

\newcommand{\pp}      {\ensuremath{pp}\xspace}

\newcommand{\W} {\ensuremath{W}\xspace}

\newcommand{\Hig} {\ensuremath{H}\xspace}

\newcommand{\ttH} {\ensuremath{\ttbar\Hig}\xspace}

\mathchardef\Upsilon="7107
\def\Y#1S{\ensuremath{\Upsilon{(#1S)}}\xspace}

\newcommand{\mt}{\ensuremath{m_{t}}\xspace}

\newcommand{\Kbar    }{\kern 0.2em\overline{\kern -0.2em K}{}\xspace}

\newcommand{\Kz      }{\ensuremath{K^0}\xspace}
\newcommand{\Kzb     }{\ensuremath{\Kbar^0}\xspace}
\newcommand{\KzKzb   }{\ensuremath{\Kz \kern -0.16em \Kzb}\xspace}
\newcommand{\Kp      }{\ensuremath{K^+}\xspace}
\newcommand{\Km      }{\ensuremath{K^-}\xspace}

\newcommand{\KpKm    }{\ensuremath{\Kp \kern -0.16em \Km}\xspace}

\newcommand\Dbar    {\kern 0.18em\overline{\kern -0.18em D}{}\xspace}

\newcommand\Bbar    {\kern 0.18em\overline{\kern -0.18em B}{}\xspace}

\newcommand\Bz      {\ensuremath{B^0}\xspace}

\newcommand\Bzb     {\ensuremath{\Bbar^0}\xspace}
\newcommand\Bu      {\ensuremath{B^+}\xspace}
\newcommand\Bub     {\ensuremath{B^-}\xspace}

\newcommand\BpBm    {\ensuremath{\Bu {\kern -0.16em \Bub}}\xspace}
\newcommand\Bs      {\ensuremath{B^0_{s}}\xspace}
\newcommand\Bsb     {\ensuremath{\Bbar^0_{s}}\xspace}

\newcommand\BzBzb   {\ensuremath{\Bz {\kern -0.16em \Bzb}}\xspace}
\newcommand\BszBszb {\ensuremath{\Bs {\kern -0.16em \Bsb}}\xspace}

\newcommand\iab{\ensuremath{\:{\rm ab}^{-1}}\xspace}
\newcommand\invfb{\ensuremath{\:{\rm fb}^{-1}}\xspace}
\newcommand\ifb{\ensuremath{\:{\rm fb}^{-1}}\xspace}

\newcommand{\tev}{\ensuremath{\,\mathrm{Te\kern -0.1em V}}\xspace}
\newcommand{\gev}{\ensuremath{\,\mathrm{Ge\kern -0.1em V}}\xspace}
\newcommand{\mev}{\ensuremath{\,\mathrm{Me\kern -0.1em V}}\xspace}

\newcommand{\GeV}{\gev}
\newcommand{\TeV}{\tev}
\newcommand\fb{\ensuremath{\,\mathrm{fb}}\xspace}
\newcommand\pb{\ensuremath{\,\mathrm{pb}}\xspace}

\newcommand{\kev}{\ensuremath{\,\mathrm{ke\kern -0.1em V}}\xspace}
\newcommand{\ev}{\ensuremath{\,\mathrm{e\kern -0.1em V}}\xspace}
\newcommand{\gevc}{\ensuremath{{\,\mathrm{Ge\kern -0.1em V\!/}c}}\xspace}
\newcommand{\mevc}{\ensuremath{{\,\mathrm{Me\kern -0.1em V\!/}c}}\xspace}
\newcommand{\gevcc}{\ensuremath{{\,\mathrm{Ge\kern -0.1em V\!/}c^2}}\xspace}
\newcommand{\mevcc}{\ensuremath{{\,\mathrm{Me\kern -0.1em V\!/}c^2}}\xspace}

\newcommand\eg{{e.g.}\xspace}

\def\@citex[#1]#2{\if@filesw\immediate\write\@auxout{\string\citation{#2}}\fi
  \@tempcnta\z@\@tempcntb\m@ne\def\@citea{}\@cite{\@for\@citeb:=#2\do
    {\@ifundefined
       {b@\@citeb}{\@citeo\@tempcntb\m@ne\@citea
        \def\@citea{,\penalty\@m\ }{\bf ?}\@warning
       {Citation `\@citeb' on page \thepage \space undefined}}%
    {\setbox\z@\hbox{\global\@tempcntc0\csname b@\@citeb\endcsname\relax}%
     \ifnum\@tempcntc=\z@ \@citeo\@tempcntb\m@ne
       \@citea\def\@citea{,\penalty\@m}
       \hbox{\csname b@\@citeb\endcsname}%
     \else
      \advance\@tempcntb\@ne
      \ifnum\@tempcntb=\@tempcntc
      \else\advance\@tempcntb\m@ne\@citeo
      \@tempcnta\@tempcntc\@tempcntb\@tempcntc\fi\fi}}\@citeo}{#1}}

\def\@citeo{\ifnum\@tempcnta>\@tempcntb\else\@citea
  \def\@citea{,\penalty\@m}%
  \ifnum\@tempcnta=\@tempcntb\the\@tempcnta\else
   {\advance\@tempcnta\@ne\ifnum\@tempcnta=\@tempcntb \else
\def\@citea{--}\fi
    \advance\@tempcnta\m@ne\the\@tempcnta\@citea\the\@tempcntb}\fi\fi}

\xspace

\begin{document}

\title{Dispelling the $\rL$ myth for the High-Luminosity LHC}

\author[a]{Alberto Belvedere,}
\author[b]{Christoph Englert,}
\author[a]{Roman Kogler,}
\author[c]{and Michael Spannowsky}

\affiliation[a]{Deutsches Elektronen-Synchrotron DESY, Notkestr. 85, 22607 Hamburg, Germany}
\affiliation[b]{School of Physics and Astronomy, University of Glasgow, Glasgow G12 8QQ, United Kingdom}
\affiliation[c]{Institute for Particle Physics Phenomenology, Department of Physics, Durham University, Durham DH1 3LE, United Kingdom}

\emailAdd{alberto.belvedere@desy.de}
\emailAdd{christoph.englert@glasgow.ac.uk}
\emailAdd{roman.kogler@desy.de}
\emailAdd{michael.spannowsky@durham.ac.uk}

\abstract{Extrapolations of sensitivity to new interactions and standard model parameters critically inform the programme at the Large Hadron Collider (LHC) and potential future collider cases. To this end, statistical considerations based on inclusive quantities and established analysis strategies typically give rise to a sensitivity scaling with the square root of the luminosity, $\sqrt{\mathcal{L}}$. This suggests only a mild sensitivity improvement for LHC's high-luminosity phase, compared to the collected data up to the year 2025. We discuss representative analyses in top quark, Higgs boson and electroweak gauge boson phenomenology and provide clear evidence that the assumption that the sensitivity in searches and measurement scales only with $\sqrt{\mathcal{L}}$ at the High-Luminosity LHC is overly conservative at best, and unrealistic in practice. As kinematic coverage enables more targeted search strategies with sufficient statistical sensitivity, employing a multitude of features in data dispels the scaling based on more inclusive selections.}

\preprint{IPPP/23/76, DESY-24-016}

\maketitle


\section{Introduction}
\label{sec:intro}
The ongoing LHC runs will define the research activities in collider phenomenology in the coming years. Currently, the LHC is in its Run-3 phase with an anticipated upgrade into the high-luminosity (HL) phase in 2029. The data from Run 2 (2016--2018) corresponds to an integrated luminosity of about 140\ifb, which will increase to an expected 300\ifb at the end of Run 3 for each multi-purpose experiment ATLAS and CMS. The HL-LHC will result in an accelerated data acquisition, where both experiments will accrue over 3000\ifb of data until the end of the LHC's lifetime. While it is impressive to increase the existing data tenfold in a relatively short amount of time, it is generally believed that the sensitivity $\mathcal{S}$ in searches for resonances and the underlying dynamics of a theory that extends the standard model (SM) scales as the square root of the integrated luminosity $\rL$, i.e. 
\begin{equation}
\mathcal{S}\simeq \frac{S}{\sqrt{B}} \simeq \rL \frac{\sigma_S}{ \sqrt{\sigma_B}},
\end{equation}
assuming LHC upgrades do not change the centre-of-mass energy $\sqrt{s}$, and therefore the cross sections of signal, $\sigma_S$, or background, $\sigma_B$, stay unchanged.
Similarly, for SM measurements with a negligible amount of background events $B$ compared to signal events $S$, the statistical uncertainty $\delta$ is obtained from the variance of a Poisson distribution and hence scales as $\delta \sim \sqrt{S} \sim \rL$. Despite a considerable increase in the LHC's luminosity, this implies only a modest sensitivity increase in searches for new physics and in measuring SM parameters. More complex is the prediction of the evolution of systematic uncertainties, originating from different sources that constitute the leading uncertainties in some cases. On this basis of a simple rescaling of the number of signal and background events with the integrated luminosity, several analyses have assessed the LHC physics potential~\cite{Dainese:2019rgk} for the HL-LHC. In these analyses, the systematic uncertainties have been estimated assuming different scenarios, reaching from unchanged uncertainties with respect to existing analyses to reducing them by either a factor of two or by \rL scaling. The HL-LHC projections contribute to a relatively gloomy outlook for the collider physics programme of the coming decades. Some of these projections and their updates continue to inform the future strategy of particle physics~\cite{EuropeanStrategyforParticlePhysicsPreparatoryGroup:2019qin,Butler:2023eah}.

However, we argue that the anticipated \rL scaling is too conservative for many relevant searches for effects beyond the SM (BSM) and SM measurements. While it is correct that the cross sections for signal and background remain essentially unchanged during the HL-LHC runs, new observables will become accessible, which will lead to a much more significant gain in sensitivity than widely projected~\cite{Azzi:2019yne, Cepeda:2019klc, CidVidal:2018eel}. Concretely, entirely new phase space regions, e.g. high transverse momentum or large invariant-mass final states, will be populated by an appreciable number of events, opening them up for tailored search and measurement strategies. Thus, novel reconstruction techniques designed for these exclusive phase space regions will significantly boost the exploitation of kinematic differences between signal and background, enhancing sensitivity in new physics searches. In addition, in any search, the experimentally measured data are tensioned with a model assumption, i.e.\ often a high-dimensional extension of the SM; accessing such new observables will help to overconstrain the parameter space of such models, thereby increasing the sensitivity far beyond the estimated \rL scaling. As the amount of data increases, new reconstruction techniques and calibrations become available, not only reducing the systematic uncertainties estimated in previous analyses but also making new measurement strategies possible. Furthermore, once sufficiently large datasets become available for constructing multi-dimensional measurement regions, the measured data can be used to constrain model parameters thus reducing modelling uncertainties. 

We showcase these observations explicitly in four example studies for representative scenarios and processes. To begin, in Sec.~\ref{sec:top}, we focus on top quark physics. Firstly, in Sec.~\ref{sec:4top}, we consider the production of four top quarks. The rich final state allows to expand the analysis to more decay channels when more data become available. The extension to channels that have not been considered in earlier versions is possible because of potent algorithms for background reduction, complemented by sideband regions that tightly constrain background yields because of better statistical precision. This showcases the prowess of novel machine learning (ML) reconstruction techniques and model constraints from sideband regions. Secondly, we turn from measuring the cross section of four top quark production to measuring one of the most relevant fundamental parameters for electroweak physics, the top quark mass. In Sec.~\ref{sec:topmass}, we use the example of fully hadronic top quark decays merged into a single jet to demonstrate how novel reconstruction algorithms and improved calibration methods greatly enhance the potential of this measurement. The measurement utilises the benefit of exclusive high-transverse momentum (\pt) phase space regions over inclusive measurements to extract additional information. We then turn towards electroweak physics in Sec.~\ref{sec:ew}. Rare processes in the SM only become accessible at large integrated luminosities. We provide a phenomenological analysis for \tWZ and \ttZ production, demonstrating how additional background reduction techniques can greatly enhance the potential of measuring the \tWZ process over the known \ttZ background. In addition, we show how a differential measurement can enhance the sensitivity to new physics interactions compared to an inclusive measurement. Finally, we show that \rL scaling is too conservative for Higgs phenomenology, too. In a fourth example, in Sec.~\ref{sec:Higgs}, we turn our attention to the leading Higgs boson production 
processes at the LHC. A comprehensive set of differential measurements can lift blind directions in the ample parameter space of new physics, encoded as Wilson coefficients of an effective field theory, breaking evidently the \rL scaling.

This representative set of examples highlights that the \rL scaling for the HL-LHC provides a too pessimistic outlook for the coming decades of collider physics. Instead, the focus on exclusive phase space regions, designated reconstruction techniques, simultaneous access to multi-dimensional measurement regions, and advanced background reduction algorithms will enable us to excel in our ability to gain a deeper understanding of the underlying dynamics of particle physics.

\section{Top physics}
\label{sec:top}
The HL-LHC is an ideal machine to study the properties of the top quark. With an integrated production cross-section of about 850 and 950\pb at $\sqrt{s}=13$ and 14\TeV, respectively~\cite{Czakon:2013goa,CMS:2023qyl}, both ATLAS and CMS will each have produced between 2.5--3.0 billion top-quark pair events by the end the HL-LHC's runtime. Because of its short lifetime with a mass close to the electroweak scale, and its characteristic decay into three fermions enabling the separation of top quark events from large QCD backgrounds, top quark studies at the HL-LHC will yield important insights into the top quark's role in the mechanism of electroweak-symmetry breaking. Thus, exhausting the HL-LHC's potential in determining all top quark properties is of fundamental importance for its success.
\subsection{Multi-top production}
\label{sec:4top}
\begin{figure*}[!t]
\centering
\includegraphics[width=0.68\textwidth]{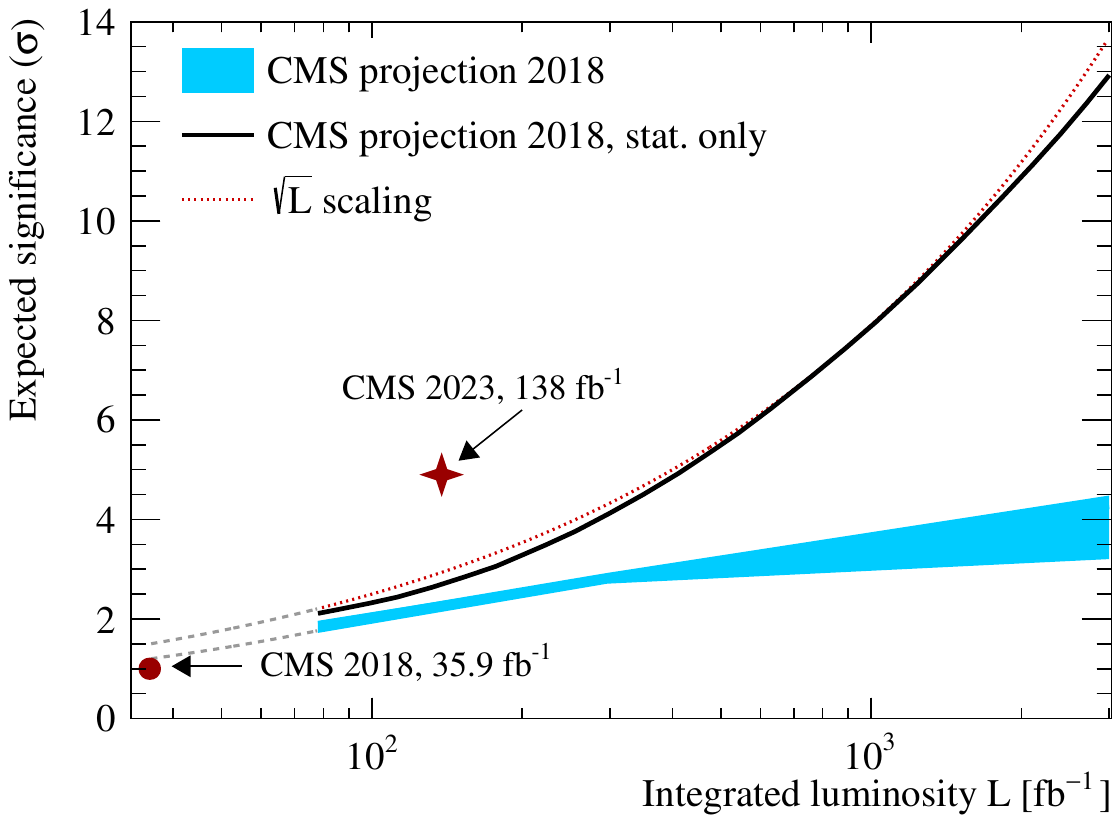}
\caption{Evolution of the expected significance in four top quark analyses by the CMS experiment.  
We compare the 2018 result based on an integrated luminosity 35.9\invfb~\cite{CMS:2017ocm} red (red filled circle), with a projection from 2018~\cite{CMS-PAS-FTR-18-031} (blue area), with the expected significance from the 2023 observation based on an integrated luminosity 138\invfb~\cite{CMS:2023ftu} (red cross). The CMS projection based on statistical uncertainties only is also shown (black solid line). The projection starts at an integrated luminosity of 78\invfb; the dashed grey lines extrapolate it with a \rL dependence down to 36\invfb. For illustration, the function $0.25 \sqrt{\mathcal{L/\!\fb}}$, is also shown (red dotted line), which describes well the expected significance obtained from the statistical-only uncertainty. 
\label{fig:4tops}}
\end{figure*}

Four top quark final states (\fourtop) are known to be sensitive to a plethora of resonant and non-resonant BSM interactions~\cite{Banelli:2020iau,Aoude:2022deh,Englert:2019zmt,Blekman:2022jag,Alvarez:2019uxp,Anisha:2023xmh}. The production of \fourtop is therefore expected to provide high sensitivity to BSM physics when the SM prediction is accurately known. The \fourtop production has a very small cross section of $13.4^{+1.0}_{-1.8}\fb$~\cite{vanBeekveld:2022hty} at 13\TeV, with an overwhelming amount of SM backgrounds, making the search for \fourtop experimentally very challenging. However, searches for \fourtop final states also demonstrate the ability of analyses to break the \rL scaling {\emph{already}}. As more data become available, increasingly exclusive selections can be achieved to combat contributing backgrounds without compromising the robustness of predictions. In Fig.~\ref{fig:4tops}, we compare the 2018 CMS expected sensitivity~\cite{CMS:2017ocm} and its extrapolation to the HL-LHC~\cite{CMS-PAS-FTR-18-031} with the recent expected sensitivity from the 2023 observation of \fourtop production~\cite{CMS:2023ftu}. The 2018 analysis uses same-sign dilepton and multilepton final states, and observes \fourtop production with an expected sensitivity of one standard deviation ($\sigma$) above the SM backgrounds, using data corresponding to 35.9\ifb of integrated luminosity~\cite{CMS:2017ocm}, shown as a filled red circle. When extrapolating this result to the HL-LHC, the cross section of \fourtop production increases by a factor of about 1.3 when increasing $\sqrt{s}$ from 13 to 14\TeV. This increase, together with an expected increase of the integrated luminosity to 78\ifb, led to a predicted sensitivity of about $2\sigma$ above the SM background in the projection, obtained by using \rL scaling~\cite{CMS-PAS-FTR-18-031}. The projection of the expected \fourtop significance for the HL-LHC is performed using two different scenarios, namely statistical uncertainties only (solid line), and three different assumptions on the systematic uncertainties (blue band). When including systematic uncertainties, the most optimistic scenario leads to an expected significance of $4.1\sigma$ with 3\iab, where the improvement from 300\ifb to the HL-LHC is only about one standard deviation. Only in the unrealistic case of statistical uncertainties only, the projection results in an observation with a significance above $5\sigma$. We argue that this pessimistic scenario is misleading because it does not consider methodical and technical improvements that can improve the sensitivity far beyond a simple reduction of systematic uncertainties in an existing analysis. This is demonstrated by a recent result by the CMS Collaboration, observing \fourtop production with an expected significance of $4.9\sigma$ using 138\ifb of 13\TeV data, shown by the star in Fig.~\ref{fig:4tops}, and an observed significance of $5.6\sigma$~\cite{CMS:2023ftu}, which is much better than anticipated from the sensitivity estimates. The analysis from 2023 updates the previous analyses in this channel~\cite{CMS:2017ocm, CMS:2019rvj} with a significantly improved lepton identification, especially at low transverse momenta. This is achieved by using Boosted Decision Trees (BDTs) to discriminate between leptons produced in the decay of charm and bottom hadrons from those produced in the decay of \W bosons. In addition, BDTs are employed to discriminate between \fourtop production and the large SM backgrounds. The analysis leverages a set of critical variables, including jet multiplicity, jet properties, and the number of jets identified to originate from \bq quarks (\bq jets), supplemented by associated kinematic variables. These multivariate analysis techniques have proven to be pivotal in isolating and studying the rare \fourtop production, achieving a significance much better than the \rL-predicted significance of $2.7\sigma$ at 138\ifb, even though this prediction is based on statistical uncertainties only, for a \fourtop cross section 1.3 times as large.\footnote{The 2018 analysis~\cite{CMS:2017ocm} was already updated in 2020 with an analysis using 137\ifb of 13\TeV data, which resulted in an expected sensitivity of $2.7\sigma$ above the SM background~\cite{CMS:2019rvj}. This result alone breaks the \rL scaling, considering that the 2018 analysis based on 35.9\ifb of integrated luminosity shows a significance of~$1\sigma$.}

A very similar analysis can be made for the search of \fourtop production with the ATLAS experiment. An ATLAS search for \fourtop final states with 36.1\ifb of 13\TeV data resulted in an expected upper limit on the production cross section of 29\fb at 95\% confidence level~\cite{ATLAS:2018alq}. The \rL scaling of this result predicted a significance of about $5\sigma$ with 300\ifb of integrated luminosity at $\sqrt{s}=14\TeV$~\cite{ATL-PHYS-PUB-2018-047}. The most recent ATLAS result in this channel achieves an expected significance of $4.3\sigma$ already with 140\ifb of 13\TeV data~\cite{ATLAS:2023ajo}, overcoming the \rL-scaling expectation. This publication reported the first observation of \fourtop production with an observed significance of $6.1\sigma$. 

In addition to the same-sign dilepton and multilepton final states, there are other channels that can be considered in the search for \fourtop production. A combination of all-hadronic, one lepton and opposite-sign dilepton events improves the expected sensitivity from $2.7\sigma$ as obtained in Ref.~\cite{CMS:2019rvj}, to $3.2\sigma$~\cite{CMS:2023zdh}. This example of \fourtop production is educative for many other processes at the LHC, where the availability of more data opens up channels that are not accessible in the first stage of data analyses and allows for a refined statistical treatment that improves the sensitivity far beyond early expectations. 

\subsection{Exclusive top quark mass measurements}
\label{sec:topmass}
Measuring the top-quark mass \mt to high accuracy is a challenge at the LHC. While top quark-pair final states can be separated fairly straightforwardly from QCD and electroweak backgrounds, measuring \mt hinges on controlling the jet-energy resolution, pileup effects from overlapping \pp-collision events, and the jet-energy scale specific to jets containing bottom quarks. Ongoing efforts to determine the top mass rely on fully leptonic~\cite{ATLAS:2014nxi,CMS:2016yys,CMS:2018fks}, semi-leptonic~\cite{ATLAS:2018fwq,CMS:2018quc} and fully hadronic~\cite{ATLAS:2017lox} top-quark decays. Traditionally, precision measurements of \mt use \pt-thresholds for the lepton, missing transverse momentum and jets as low as possible, for minimal statistical uncertainties due to the cross section-enhanced production of top quarks with small transverse momentum. However, boosted top quark final states~\cite{Larkoski:2017jix, Asquith:2018igt, Kogler:2021kkw} provide several advantages in ameliorating experimental challenges in reconstructing hadronically decaying top quarks~\cite{Plehn:2011tg}, even to a degree where these final states can improve on the sensitivity in the cleaner final states with leptonic top decays~\cite{Plehn:2010st,Plehn:2011tf}. Concretely, one trades the smaller cross section of a boosted final state against the larger cross section of a fully hadronic top decay compared to a leptonic top decay, while benefitting from the reconstruction advantage by finding all the hadronic decay products in a relatively small confined area of a detector, i.e.\ inside a large jet. A measurement of \mt in this boosted topology is affected by complementary systematic uncertainties to measurements from low-\pt final states. In addition, it allows to probe \mt at energy scales much higher than previously reached and can help to resolve ambiguities in relating \mt measurements to its expression in well-defined theoretical schemes~\cite{Hoang:2017kmk, Hoang:2020iah, Butenschoen:2016lpz, Dehnadi:2023msm}. 

\begin{figure*}[t]
\centering
\includegraphics[width=0.68\textwidth]{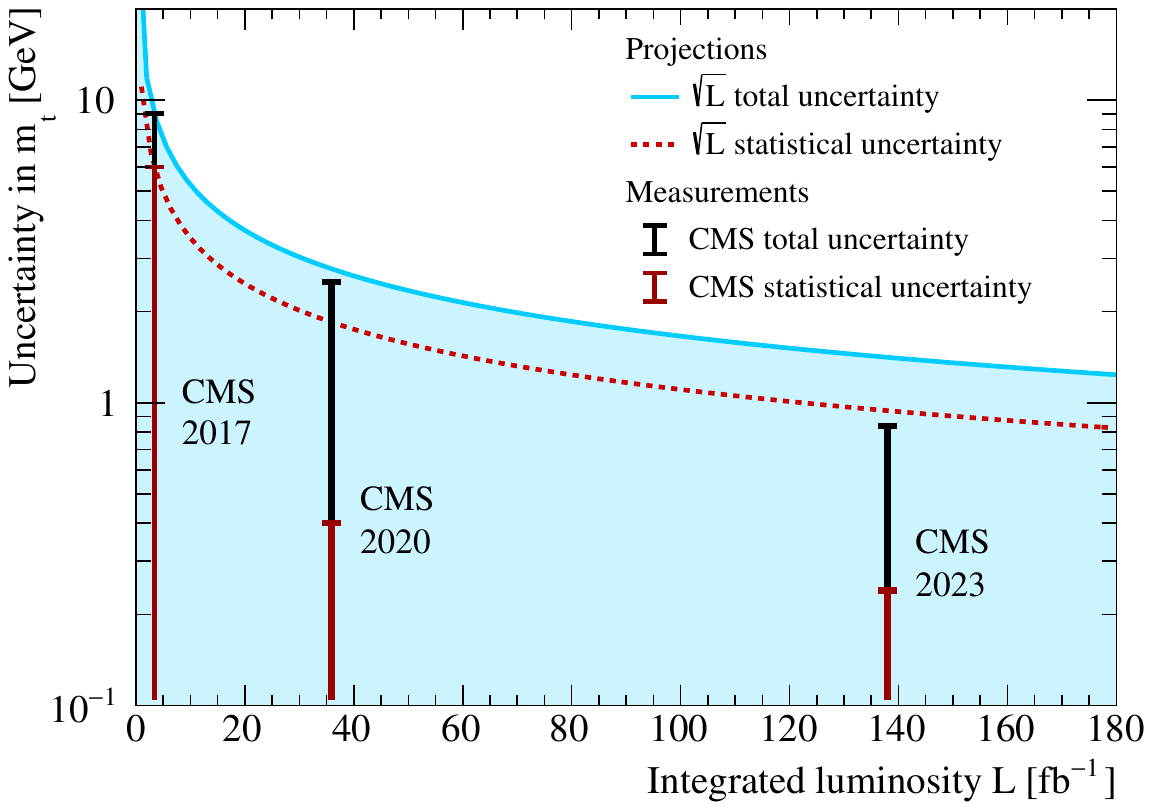}
\caption{Projected evolution of the total (blue solid line) and statistical (red dashed line) uncertainties when measuring \mt from the jet mass in hadronic decays of boosted top quarks by the CMS Collaboration. The projected uncertainties are obtained using \rL scaling of the 8\TeV measurement from 2017~\cite{CMS:2017pcy}, for an effective integrated luminosity of 3.4\invfb corresponding to an equivalent measurement at 13\TeV (see main text for details). The projected uncertainties are compared to CMS measurements from 2020~\cite{CMS:2019fak} and 2023~\cite{CMS:2022kqg}, using 13\TeV data corresponding to integrated luminosities of 35.9 and 138\invfb, respectively.
\label{fig:tmass}
}
\end{figure*}
The first determination of \mt from the measured cross section as a function of the jet mass was performed by the CMS Collaboration in 2017 with 8\TeV \pp data corresponding to an integrated luminosity of 19.7\invfb~\cite{CMS:2017pcy}. The hadronic top quark decays were reconstructed using a single large-radius jet with $\pt>400\GeV$. Then, the differential top quark pair production cross section was unfolded as a function of the jet mass to the particle level, which was used to extract \mt. We use this measurement to predict the evolution of statistical and systematic uncertainties with \rL scaling. This measurement has been performed at $\sqrt{s}=8\TeV$, but we will compare it to measurements based on 13\TeV data. To obtain comparable sensitivities, we scale the integrated luminosity of the 8\TeV measurement such that the predicted number of events is the same as for a measurement at 13\TeV. This is achieved by taking the cross-section ratio between 8 and 13\TeV for the phase space of this measurement, resulting in an effective integrated luminosity of 3.4\invfb at $\sqrt{s}=13\TeV$. In other words, we expect a measurement at 13\TeV with an integrated luminosity of 3.4\invfb to have the same statistical and systematic uncertainties as the 8\TeV measurement. 
For the prediction of the sensitivity, we scale not only the statistical but also the systematic uncertainty with \rL, which is arguably too optimistic, and leads to a decrease of the total uncertainty proportional to \rL as a function of the integrated luminosity, as shown by the blue region in Fig.~\ref{fig:tmass}. The next measurement of \mt using the jet mass was published in 2020 on 13\TeV data with an integrated luminosity of 35.9\ifb~\cite{CMS:2019fak}, shown as a second bar in Fig.~\ref{fig:tmass}. The statistical uncertainty of 0.4\GeV in \mt is much smaller than the projected 1.8\GeV obtained from the \rL scaling. This has been achieved by an improved jet reconstruction using the XCone algorithm~\cite{Stewart:2015waa} with a two-step clustering~\cite{Thaler:2015xaa}. This improved the width of the lineshape in the jet mass distribution as well as the experimental resolution, leading to a much larger event count in the peak region and therefore a reduced statistical uncertainty. Even the total uncertainty is smaller than the optimistic \rL projection, because of a more precise calibration of the jet mass and a largely reduced susceptibility to pileup. The most recent measurement by the CMS Collaboration was published in 2023, using the full Run-2 dataset corresponding to an integrated luminosity of 138\invfb~\cite{CMS:2022kqg}. For this measurement, the jet mass scale was calibrated using the hadronic $W$ boson decay within the large-radius jet, and uncertainties in the modelling of the final state radiation were reduced with the help of an auxiliary measurement of angular correlations in the jet substructure. This approach led to a significant increase in precision, culminating in $\mt = 173.06 \pm 0.84\GeV$~\cite{CMS:2022kqg}. We note that the precision is much better than the optimistic \rL projection of the total uncertainty obtained from the 8\TeV measurement, which gives 1.4\GeV. It is even better than the \rL scaling of the statistical uncertainty only, which results in a projected uncertainty of 0.94\GeV. This example prominently highlights the improvements in precision possible when developing advanced data analysis strategies, information from auxiliary measurements, and experimental calibration methods.

\section{Electroweak gauge boson interactions}
\label{sec:ew}
\subsection{Rare electroweak processes: A \texorpdfstring{$\tWZ$}{tWZ} case study}
\label{sec:twz}
Processes involving top quarks ($t$) and EW gauge bosons ($W$, $Z$, $\gamma$) reveal high sensitivity to BSM effects~\cite{Maltoni:2019aot} and allow to probe the top-EW interaction, which is only poorly constrained by experimental data~\cite{Hartland:2019bjb, Ellis:2020unq}. The study of the process $\pp \to \tWZ$ is crucial for probing the weak couplings of the top quark. This process is particularly sensitive to unitarity-violating behaviour due to modified electroweak interactions, setting it apart from other electroweak top production processes. Unlike the dominant QCD-induced production modes like \ttZ, which primarily experience rate rescalings from operators, \tWZ offers a unique perspective due to its sensitivity to the weak interactions of top and bottom quarks, and the self-interactions of $\SU(2)$ gauge bosons. This sensitivity is not as pronounced in related processes such as $\tWj$ and $\tZj$, making $\tWZ$ a more distinct and effective probe. Additionally, $\tWZ$ is not influenced by top quark four fermion operators at the tree level, further enhancing its significance as a complementary tool for exploring new interactions in the top quark sector~\cite{Faham:2021zet}.  

The first evidence of the standard model production of a top quark in association with a $W$ and a $Z$ boson in multi-lepton final states was reported by CMS~\cite{CMS:2023krq}, using data from 2016 to 2018, with an integrated luminosity of $138\ifb$. The measured cross section was found to be $354 \pm 54 \mathrm{(stat)} \pm 95 \mathrm{(syst)}$\fb, with an observed significance of $3.4$ standard deviations, compared to an expected significance of $1.4$ standard deviations. 

As a rare electroweak process that probes a range of relevant top-related interactions, the sensitivity of \tWZ production to SMEFT operators at the HL-LHC has been studied in Ref.~\cite{Keaveney:2021dfa}. Couplings that are probed in the top sector are particularly relevant in scenarios of partial top compositenss~\cite{Brown:2020uwk} or general vector-like quark extensions~\cite{Alves:2023ufm}. To highlight the potential of this rare process going forward with detailed kinematic information becoming increasingly accessible, we consider the $\mathcal{O}_{tZ}$ and the $\mathcal{O}_{\phi Q}^3$ operators as defined in the SMEFT@NLO~\cite{Degrande_2021} model with
\begin{equation}
\mathcal{O}_{tZ} = -\sin\theta_W\mathcal{O}_{tB} + \cos\theta_W\mathcal{O}_{tW} \,.
\end{equation}
The operators on the right-hand side refer to the Warsaw basis convention~\cite{Grzadkowski_2010} and $\theta_W$ is the weak mixing angle. Although not an exhaustive list, these couplings provide relevant and representative deformations from the expected $\tWZ$ SM outcome, and the $\tWZ$ process is particularly sensitive to these types of operators. The similarity of the final state to the $\ttZ$ process complicates the experimental identification of $\tWZ$ significantly. For this reason, we study the effect of the $\mathcal{O}_{tZ}$ and $\mathcal{O}_{\phi Q}^3$ operators for both processes simultaneously. 

To simulate the two processes we use Madgraph5 aMC@NLO~\cite{Alwall_2014}, while the EFT effects are taken into account using the SMEFT@NLO model. We generate $\pp \to \tWZ$ and $\pp \to \ttZ$ at $\sqrt{s} = 13\TeV$ under the SM assumption, where the simulation of \tWZ makes use of the diagram removal technique~\cite{Frixione:2008yi, Demartin:2016axk}. Then the events are reweighted to different BSM points by simulating the effects of the previously mentioned EFT operators for different values of the Wilson coefficients using the reweighting method~\cite{Mattelaer_2016}. The events are generated at next-to-leading order in QCD, while the parton shower is modelled using PYTHIA8~\cite{Sj_strand_2015}. The new physics scale is chosen to be $\Lambda=1\TeV$ and including quadratic terms $\sim \Lambda^{-4}$ to the analysis at this point to highlight the power of high-energy search regions~\cite{Faham:2021zet}. 

\begin{figure*}[tb]
\centering
\includegraphics[height=0.55\textwidth]{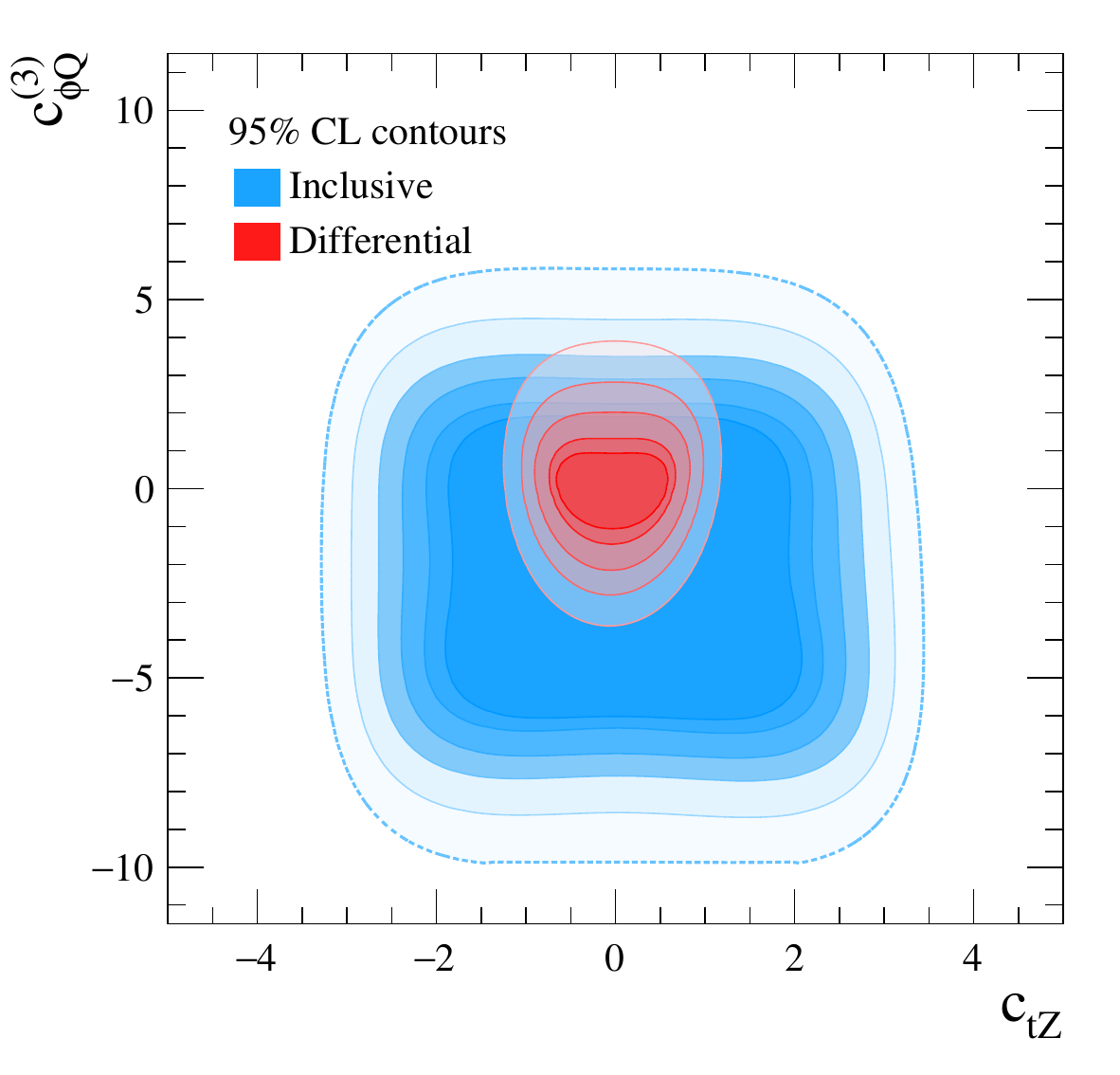}
\includegraphics[height=0.55\textwidth, trim={2.5cm 7cm 5cm 1cm}, clip]{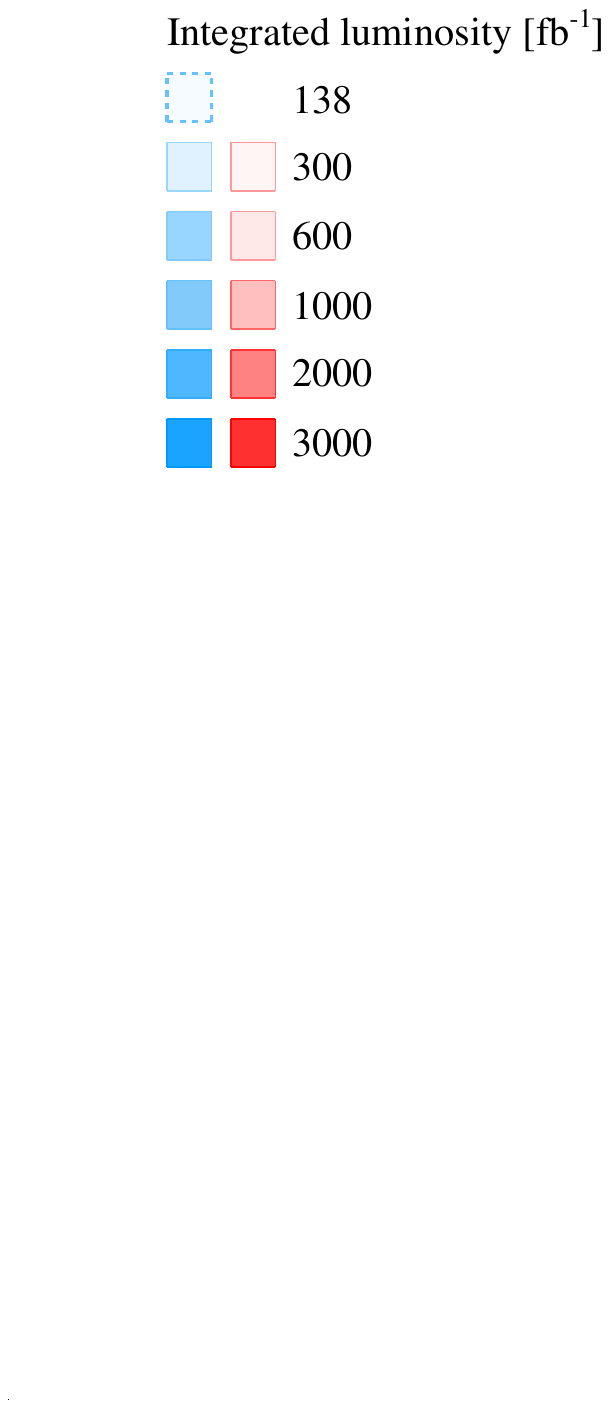}
\caption{Constraints from a fit to \ttZ and \tWZ simulated data, as described in the text, on the SMEFT operators $\mathcal{O}_{tZ}$ and $\mathcal{O}_{\phi{}Q}^{(3)}$, for $\Lambda=1\TeV$. The blue-shaded area shows constraints from inclusive measurements, while the red-shaded area refers to constraints including differential $p_{\text{T},Z}$ measurements. The outermost blue area bounded by the dashed line corresponds to expected constraints from the CMS Run 2 measurement of \tWZ production~\cite{CMS:2023krq}.
\label{fig:ctzcpq3}}
\end{figure*}

We simulate the \ttZ{}+\tWZ measurement by including realistic efficiencies and acceptances from the CMS experiment.
Backgrounds from dibsoson and $\ttbar{}+X$ production, as well as backgrounds from misidentified leptons (non-prompt backgrounds), are estimated from the recent CMS analysis~\cite{CMS:2023krq}. Our analysis considers three- and four-lepton final states. For each final state, the events are separated into a \tWZ signal region (SR) and a \ttZ control region (CR). 
The event yields are obtained using the CMS reconstruction and identification efficiencies, 
most importantly those for electrons~\cite{CMS:2020uim} and muons~\cite{CMS:2018rym}. The signal acceptances in the SR and CR, as well as systematic uncertainties, are estimated from Ref.~\cite{CMS:2023krq}. With this setup, we reproduce the CMS results in terms of signal yields and signal strengths for \ttZ and \tWZ within about 10\%.

Interpreting these results in the context of the SMEFT, the results are shown in Fig.~\ref{fig:ctzcpq3}. The dashed line corresponds to the expected limits at 95\% confidence level on $c_{tZ}$ and $c_{\phi Q}^3$ from the CMS analysis with an integrated luminosity of $138\ifb$. Four regions enter this analysis, the SRs and CRs of the three- and four-lepton final states. Subsequent blue shading corresponds to increasing the luminosity using the 138\ifb results as the baseline. Besides the higher event counts, the systematic uncertainties are also reduced following an approximate \rL scaling. Given the nature of the included coupling modifications, the inclusive selection's sensitivity plateaus when uncertainties at low momentum transfers dwarf the relative deviation from the SM. This process therefore is a prime example of the $\rL$ scaling being badly broken by the nature of the physics that characterises the BSM scenario: Already at 300\ifb we expect to have enough data to populate a differential fit of these Wilson coefficients, thereby avoiding the sensitivity loss of the inclusive selection. The differential analysis is performed in the distribution of $Z$ boson \pt, with a bin size much larger than the experimental resolution such that resolution effects can be neglected. Increasing the luminosity then gains further statistical and likely systematic control, improving the separation between \ttZ and \tWZ, thus enabling much tighter constraints than the $\rL$ scaling of the inclusive selection could provide (red shading in Fig.~\ref{fig:ctzcpq3}). It is worth noting that the constraints on $c_{tZ}$ are comparably weak in marginalised fits so far, see e.g.\ Ref.~\cite{Ethier:2021bye}. A differential measurement of the combined \ttZ{}+\tWZ processes can alleviate this and is likely to dramatically improve the sensitivity to this operator.

\section{Higgs physics}
\label{sec:Higgs}
\subsection{Higgs property measurements}
\label{sec:eft}
Effective Field Theory has become the consensus for reporting and parametrising sensitivities in the electroweak sector, particularly concerning the Higgs boson. Deviations from the Higgs boson's SM phenomenology are parametrised by effective operators that reminisce the existence of BSM physics at higher mass scales. The large number of operators~\cite{Grzadkowski_2010} results in the same number of unconstrained Wilson coefficients. Because some of the operators exhibit correlated effects on measurable quantities, a parametrisation in the Wilson coefficients results in phenomenologically insensitive directions in the high-dimensional parameter space when constraints are derived from data. A famous example of this is gluon-fusion Higgs production, which leads to a blind direction 
$c_{\Phi G} = - {\frac{ \alpha_s }{ 12 \pi y_t} } c_{t\Phi}$. 
The effective interaction between the Higgs field and the gluon can not be distinguished from the effective interaction between the Higgs field and the top quark, because of the loop-induced production of the Higgs boson.
The blind direction of inclusive measurements is broken when the Compton wavelength of the top quark is experimentally resolved, \eg through \Hig{}+jet and \ttH production. The available differential information from these processes directly breaks the $\rL$ scaling that enhances sensitivity but never resolves the degeneracy present in the inclusive rate~\cite{Grojean:2013nya, Schlaffer:2014osa}. As shown in Ref.~\cite{Englert:2017aqb} for example, this discrimination is stable against expected theoretical uncertainties.

\begin{figure*}[!t]
\centering
\includegraphics[height=0.55\textwidth]{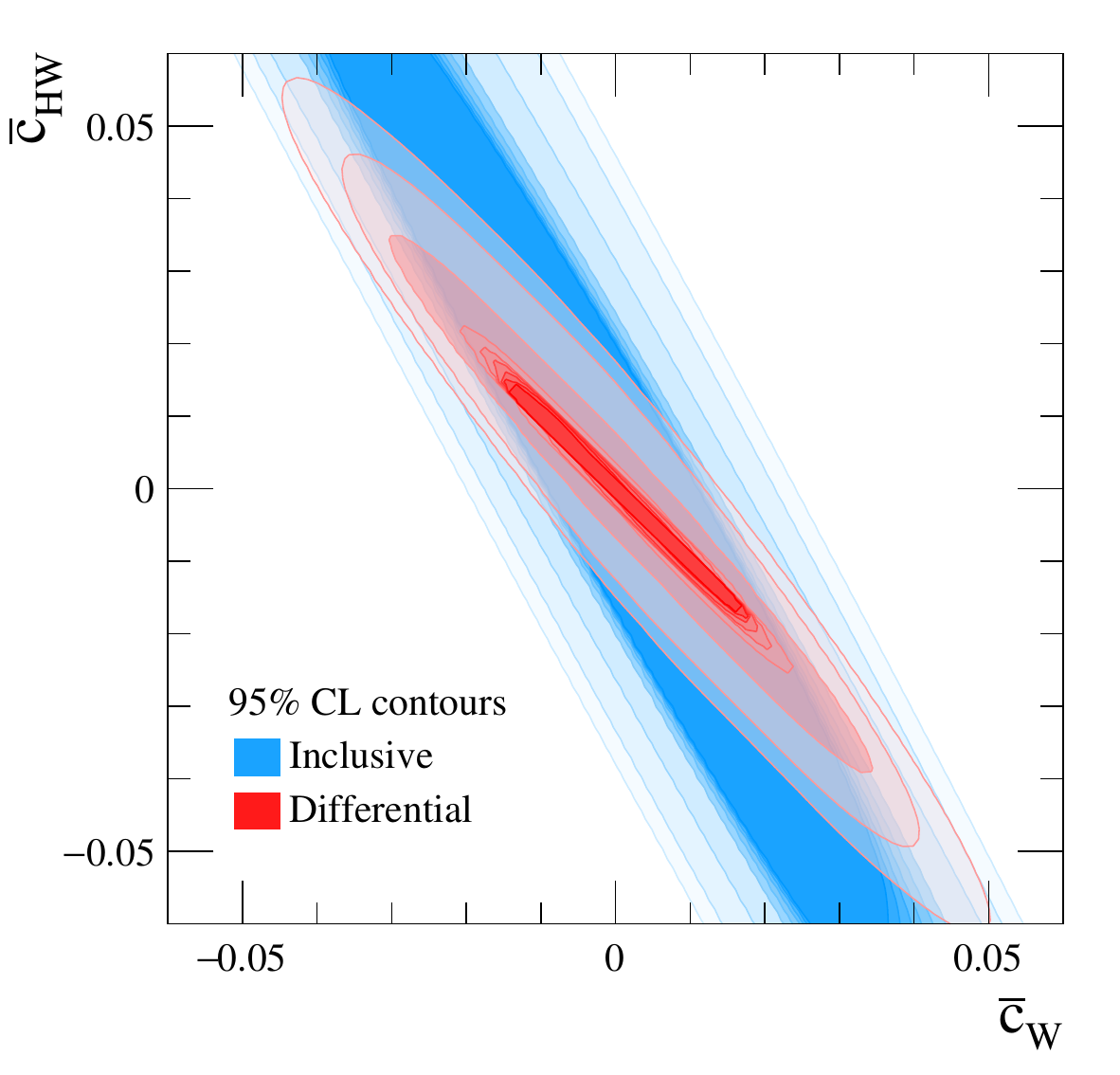}
\includegraphics[height=0.55\textwidth, trim={2.5cm 7cm 5cm 1cm}, clip]{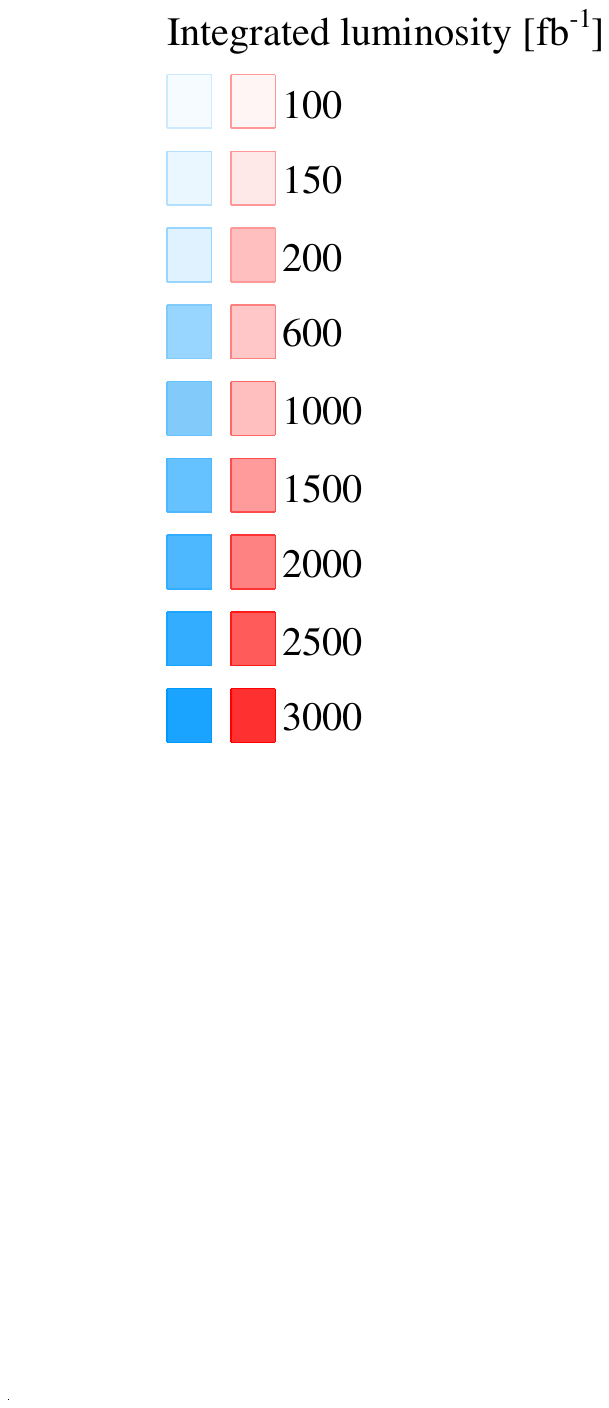}
\caption{Constraints from a global fit, as described in the text, on the SILH-basis operators $\mathcal{O}_W$ and $\mathcal{O}_{HW}$. The blue-shaded area shows constraints from inclusive measurements, while the red-shaded area refers to constraints including differential $p_{\text{T},H}$ measurements. Evaluating these constraints at various integrated luminosities, ranging from 100\ifb to 3000\ifb highlights that differential measurements are needed to resolve the blind direction residual to inclusive measurements, thereby improving in the direction $c_W \simeq -c_{HW}$ over any anticipated $\rL$ scaling.
\label{fig:cgchw}}
\end{figure*}

A similar yet slightly more dramatic behaviour is resolved when non-trivial momentum dependencies are present. This is particularly highlighted by the interplay of the $O_W$ and $O_{HW}$ operators in the SILH convention \cite{Giudice:2007fh} that parametrises non-SM momentum dependencies of the Higgs and gauge bosons \cite{Franceschini:2017xkh, Banerjee:2019twi}. 

We follow the analysis in Ref.~\cite{Englert:2015hrx} and perform a global fit, including all significant Higgs boson production and decay modes. To calculate the Higgs boson yields, we rely on the narrow-width approximation
\begin{equation}
\sigma(pp \to (H \to YY)+X)= \sigma(\pp \to H+X) \mathrm{BR}(H\to YY),
\end{equation}
where $X$ represents any associated reconstructed object, i.e. jets, top quarks or gauge bosons. To deform the Standard Model, we include the 8 operators $\bar{c}_H$, $\bar{c}_{u,3}$, $\bar{c}_{d,3}$, $\bar{c}_W$, $\bar{c}_{HW}$, $\bar{c}_{HB}$, $\bar{c}_\gamma$ and $\bar{c}_g$. We restrain the analysis to genuine dimension six effects that arise from the interference of the dimension six amplitude with the Standard Model, i.e. 
\begin{equation}
|\mathcal{M}|^2 = |\mathcal{M}_{\mathrm{SM}}|^2 + 2~\mathrm{Re} \left \{ \mathcal{M}_\mathrm{SM} \mathcal{M}^*_{d=6} \right \} + \mathcal{O}(1/\Lambda^4).
\end{equation}
Including Higgs production modes, where the Higgs boson is produced in association with jets, top quarks or gauge bosons, results in a finite transverse momentum distribution. Thus, the fit can explore the sensitivity of Higgs boson measurements in exclusive phase space regions. We bin the Higgs boson's transverse momentum in each production channel in five 100 GeV bins, from 0 to 500 GeV. To tension the theoretical predictions with experimental data, we rely on measurements obtained by ATLAS and CMS and their respective systematic and theoretical uncertainty projections. Concretely, we include unfolded $p_{\text{T},H}$ distributions for the production processes $pp \to H$, $pp \to H+j$, $pp \to H+2j$, $pp \to \bar{t}tH$ and $pp \to VH$. Decay modes of interest, included in the fit, are $H \to \bar{b}b$, $H \to \gamma \gamma$, $H \to \tau^+ \tau^-$, $H \to 4l$, $H \to 2l 2\nu$, $H \to Z\gamma$ and $H \to \mu^+ \mu^-$. For bins to be included in the fit, we require
\begin{equation}
N_{\mathrm{events}} = \epsilon_p \epsilon_d \sigma(H+X) \mathrm{BR}(H\to YY) \mathcal{L} \gtrsim 5~,
\label{eq:nrev}
\end{equation}
where $\epsilon_p$ and $\epsilon_d$ refer to the reconstruction efficiencies specific to the respective production and decay modes. A detailed description of efficiencies, acceptances, and systematic uncertainties is given in Ref.~\cite{Englert:2015hrx}. Thus, depending on the integrated luminosity $\mathcal{L}$, more or less independent measurements are included in this global fit. For 300\ifb we find 88 and for 3000\ifb 123 measurements satisfying Eq.~\eqref{eq:nrev}. Thus, this already showcases the growing amount of information from exclusive phase space regions with the increase of integrated luminosities. 

A comprehensive analysis for all Wilson coefficients can be found in \cite{Englert:2015hrx, Englert:2017aqb}. In Fig.~\ref{fig:cgchw}, we visualise the constraints obtained for $O_W$ and $O_{HW}$, while marginalising over all the other operators. The blue contour shows the inclusive measurements, i.e. only considering overall rates without considering the differential binning of $p_{\text{T},H}$. The shading refers to different integrated luminosities. The constraints considering the differential distribution of the Higgs boson are shown in red. The inclusive measurement can never resolve the inclusive blind direction for $c_W \simeq -c_{HW}$, evidenced in Fig.~\ref{fig:cgchw}. Differential measurements that access the more exclusive phase space region significantly enhance these directions' sensitivity and discrimination beyond what a $\rL$ scaling forecasts.

\section{Summary}
Extrapolations of sensitivity estimates are crucial in shaping the particle physics roadmaps and phenomenological programmes. Overly pessimistic expectations, albeit constituting a conservative point of reference, can therefore be detrimental to experimental as well as theoretical progress. Projections based on existing analyses give sensitivity improvements that scale with the square root of the luminosity ($\rL$)~\cite{Dainese:2019rgk}. While these projections are conceptually correct, the failure to include important aspects of future analyses can lead to a severe underestimation of the achievable sensitivity. In this work, we challenge the longstanding assumption that sensitivity in particle physics experiments, particularly at the High-Luminosity Large Hadron Collider (HL-LHC), scales with $\rL$. Our analysis of representative examples demonstrates that this $\rL$ scaling is exceedingly conservative, especially in the context of the HL-LHC's advanced capabilities in utilising exclusive final states and advanced reconstruction methods.

Concretely, we focused on three key areas: top quark physics, rare electroweak processes, and Higgs property measurements. These examples reveal that more differential measurements, the study of rare processes not experimentally accessible so far, increasingly refined search strategies, and advanced analysis techniques substantially enhance the experimental sensitivity, surpassing the traditional $\rL$-scaling predictions. In the realm of top quark physics, our findings indicate that the sensitivity, not only to searches for new physics in large invariant-mass final states but also in the measurement of fundamental standard model parameters, can increase significantly, providing deeper insights into the top quark's properties and interactions. Our investigations into rare electroweak processes have evidenced the potential for discoveries beyond the standard model. The enhanced sensitivity in these processes could lead to observations of new phenomena, offering a window into physics beyond our current theoretical framework. Furthermore, for Higgs property measurements, the application of new analysis methodologies has shown potential for more precise determinations of Higgs boson characteristics, a cornerstone for understanding the standard model of particle physics and beyond.

Albeit these are only examples chosen to illustrate the impact of improved and adapted data analysis strategies, they representatively demonstrate, for a wide range of applications that the success of the HL-LHC - and therefore the entire high-energy physics programme - may well hinge on the use of novel reconstruction techniques, \eg jet substructure observables, machine-learning and matrix-element methods, and the focus on signal-rich exclusive phase space regions. This poses well-known challenges for the experimental community in accessing such information through advances in tracking, calibration and particle reconstruction, for example. Furthermore, theoretical progress is crucial for the reduction of modelling and theory uncertainties, which are limiting the sensitivity of high-precision measurements already today. Thus, a wide effort from the experimental and theoretical particle physics communities is dedicated to the development of novel tools, calculations, and advanced data analysis methods, already showing a breaking of the \rL scaling in present analyses. 

Therefore, the implications of these findings suggest a change in how sensitivity is estimated for future collider experiments, by broadening these studies with unexplored final states, more differential measurements, and modern analysis techniques as more data becomes available. Our research indicates that the HL-LHC could be significantly more potent in probing the fundamental aspects of particle physics than previously anticipated. Thus, the findings dispel the myth of $\rL$ scaling and call for reevaluating experimental strategies and data analysis techniques, encouraging the scientific community to look beyond conventional assumptions and explore the full potential of the HL-LHC.

\section*{Acknowledgments}
A.B. and R.K. are supported by the Helmholtz Association under the contract W2/W3-123. C.E. is supported by the UK Science and Technology Facilities Council (STFC) under grant ST/X000605/1 and the Leverhulme Trust under Research Project Grant RPG-2021-031. M.S. is supported by the STFC under grant ST/P001246/1.

\bibliographystyle{JHEP_mod}
\bibliography{references}

\end{document}